\documentclass[5p,times,authoryear]{elsarticle}
\usepackage{graphicx,longtable}
\usepackage{natbib}
\usepackage{amssymb}
\usepackage{amsmath}
\usepackage{amsthm}
\usepackage{pdflscape}

\journal{New Astronomy}

\newcommand{\fdg}{$^{\circ}_{.}}

\def\gsim{\;\rlap{\lower 2.5pt\hbox{$\sim$}}\raise 1.5pt\hbox{$>$}\;}
\def\lsim{\;\rlap{\lower 2.5pt\hbo time lag between the formation x{$\sim$}}\raise 1.5pt\hbox{$<$}\;}
\def\la{\mathrel{\hbox{\rlap{\hbox{\lower4pt\hbox{$\sim$}}}\hbox{$<$}}}}
\def\ga{\mathrel{\hbox{\rlap{\hbox{\lower4pt\hbox{$\sim$}}}\hbox{$>$}}}}

\def\arcmin{\hbox{$^\prime$}}
\def\arcsec{\hbox{$^{\prime\prime}$}}

\def\fdg{\hbox{$.\!\!^\circ$}}
\def\farcm{\hbox{$.\mkern-4mu^\prime$}}

\def\ion#1#2{#1$\;${\small\rm{#2}}\relax}

\begin{document}
\begin{frontmatter}
\title{Star formation toward the H~II region IRAS\,10427-6032}
\author[bits]{Kaushar Vaidya\corref{cor1}}
\ead{kaushar@pilani.bits-pilani.ac.in}
\cortext[cor1]{Corresponding author}
\author[bits,eso]{Souradeep Bhattacharya}
\author[bits,ua]{Vatsal Panwar}
\author[ncu]{Manash R. Samal}
\author[ncu]{Wen-Ping Chen}
\author[tifr]{Devendra K. Ojha}
\address[bits]{Department of Physics, Birla Institute of Technology and Science - Pilani, 333031, Rajasthan, India}
\address[eso]{European Southern Observatory, Karl-Schwarzschild-Str. 2, 85748, Garching, Germany}
\address[ua]{Anton Pannekoek Institute for Astronomy, University of Amsterdam, Science Park 904, Amsterdam, Netherlands}
\address[ncu]{Graduate Institute of Astronomy, National Central University, 300 Jhongda Road, Jhongli 32001, Taiwan}
\address[tifr]{Department of Astronomy \& Astrophysics, Tata Institute of fundamental Research, Colaba, Mumbai, 400005, India}

\begin{abstract}
The formation and properties of star clusters formed at the edges of the \ion{H}{II} regions are poorly known.  In this paper, we study stellar content, physical conditions, and star formation processes around a relatively unknown young \ion{H}{II} region IRAS\,10427-6032, located in the southern outskirts of the Carina Nebula.  We study this region by making use of the near infrared (near-IR) data from VISTA, mid infrared (mid-IR) from {\it Spitzer} and WISE, far infrared (far-IR) from ${\it Herschel}$, sub-mm from ATLASGAL, and 843 MHz radio-continuum data.  Using multi-band photometry, we identify a total of 5 Class I and 29 Class II young stellar object (YSO) candidates, most of which newly identified, in the 5$\arcmin \times$ 5$\arcmin$ region centered on the IRAS source position.  Modeling of the spectral energy distribution for selected YSO candidates using the radiative transfer models shows that most of these candidates are intermediate mass YSOs in their early evolutionary stages. 
A majority of the YSO candidates  are found to be coincident with the cold dense clump at the western rim of the \ion{H}{II} region.  Lyman continuum luminosity calculation using radio emission indicates the spectral type of the ionizing source to be earlier than B0.5--B1.  We identified a candidate massive star possibly responsible     
for the \ion{H}{II} region with an estimated spectral type B0--B0.5. The temperature and column density maps of the region constructed by performing pixel-wise modified blackbody fits to the thermal dust emission using the far-IR data from the ${\it Herschel}$ show a high column density shell-like morphology around the \ion{H}{II} region, and low column density (0.6 $\times$ 10$^{22}$ cm$^{-2}$) and high temperature ($\sim$21 K) matter within the \ion{H}{II} region.  Based on the morphology of the region in the ionized and the molecular gas, and the comparison between the estimated timescales of the \ion{H}{II} region and the YSO candidates in the clump, we argue that the enhanced star-formation at the western rim of the \ion{H}{II} region is likely due to compression by the ionized gas.
\end{abstract}


\end{frontmatter}

\section{Introduction}

Massive stars have dramatic influence on their surroundings.  Due to their strong stellar winds and ionizing flux, they create bubbles/\ion{H}{II} regions which are routinely detected in mid infrared (mid-IR) wavelengths \citep{churchwell06}.  These bubbles show spatially coincident emission at mid-IR wavelengths such as {\it Spitzer} MIPS 24 $\mu$m, arising from heated dust grains, and in radio continuum such as 20 cm, due to ionized hydrogen in the bubble interiors \citep{deharveng10}.  The bubble rims, on the other hand, are defined by the emission due to polycylic aromatic hydrocarbons (PAHs) visible in certain mid-IR wavelengths including {\it Spitzer} IRAC 8.0 $\mu$m or WISE 12 $\mu$m \citep{churchwell06,deharveng10,kendrew16}.  A strong correlation is found to exist between mid-IR bubbles and cold dense clumps in which star formation is likely to occur \citep{kendrew16}.  An expanding \ion{H}{II} region may trigger star formation via the radiative-driven implosion (RDI) mechanism \citep{bertoldi89} or the collect-and-collapse (C\&C) process \citep{elmegreen77}. 

Mid-IR bubbles were studied by \citet{churchwell06} using the {\it Spitzer} survey, Galactic Legacy Infrared Mid-Plane Survey Extraordinaire (GLIMPSE).  They found that one-fourth of the bubbles in their sample had broken morphology which they attributed to lower density of the ambient interstellar medium (ISM) and/or higher ionizing photon flux in the open directions.  Whereas \citet{churchwell06} had found only 25\% of bubbles associated with \ion{H}{II} regions, \citet{deharveng10} found that as many as 86\% of bubbles enclose \ion{H}{II} regions. Moreover, they found that 40\% of the bubbles were surrounded by cold dust detected at 870 $\mu$m, whereas 28\% contained interacting condensations.  More recently, \citet{kendrew16} examined cold dense clumps detected by the ATLASGAL in and around the inner Galactic plane under the Milkyway Galaxy Project.  In their comprehensive study, they found that $\sim$48\% of the cold dense clumps are located in close proximity of bubbles, and among them $\sim$25\% appear projected toward bubble rims.  As the star-forming clouds are often fractal and clumpy, an investigation of which mechanism dominates star formation at the boarders of \ion{H}{II} regions requires understanding of the physical connection and interaction of the bubbles/\ion{H}{II} regions with the cold ISM, and their association with stellar/protostellar content and the timescales involved.

IRAS\,10427-6032 was first studied by \citet{kerber00} along with six other planetary nebula (PN) candidates.  On the basis of imaging and spectroscopic observations, \citet{kerber00} concluded that IRAS\,10427-6032 is an \ion{H}{II} region, rather than a PN.  The flux densities of IRAS\,10427-6032 as measured by IRAS are 3.24(0.16), 8.47(0.42), 92.8(1.29), and 144(2.16) Jy, in 12, 25, 60, and 100 $\mu$m, respectively, where the numbers in parenthesis are the errors in flux densities. Recently, it was identified  by \citet{anderson14} as an \ion{H}{II} region based on the all sky mid-IR data from the WISE. It is located at the southern edge of the Carina Nebula, a highly complex and massive star-forming region of our Galaxy.  The Carina Nebula is well-known for its extreme stellar content including 70 known O stars, 127 B0--B3 stars, 3 Wolf-Rayet stars, and the prototypical luminous blue variable $\eta$ Carina \citep{walborn95}.  It has also been recognized as a prolific stellar nursery \citep{smith06}.  Numerous examples of ongoing star formation are found throughout the Nebula despite the ``hostile'' environment, particularly so in the southern pillar of the Nebula \citep{megeath96,smith10,sanchawala07a,rathborne02}.  Using a large 6.7 square degree deep near-IR imaging survey of the Nebula -- the VISTA Carina Nebula Survey (VCNS) -- \citet{preibisch14a} discovered several previously unknown embedded clusters/groups of YSO candidates \citep{zeidler16}.  One such group, J104437.6-604756, is found near the southern edge of the Nebula, at about 1\fdg3 from the massive star $\eta$ Carina and close to ($\sim$15${\arcsec}$) IRAS\,10427-6032. The near-IR images show a presence of a faint cluster here.    

We found that IRAS 10427-6032 features a broken bubble.  Moreover, a compact (semi-major axis $\sim$11${\arcsec}$, semi-minor axis $\sim$5${\arcsec}$) and moderately bright (integrated flux of 3.11 Jy) cold dense clump detected by ATLASGAL,  AGAL288.069-01.645, is located at an angular distance of $\sim$15${\arcsec}$ from the IRAS source \citep{contreras13}.  Figure~1 shows the 2${\arcmin} \times$2${\arcmin}$ field around IRAS\,10427-6032 -- {\it Spitzer} 4.5 $\mu$m image in 1a, VISTA $K_s$ band in 1b, and WISE RGB image (4.6 $\mu$m in blue, 12 $\mu$m in green, and 22 $\mu$m in red) in 1c.  With the morphology of a broken bubble, presence of an \ion{H}{II} region, and cold dust condensation detected at 870 $\mu$m, it is an interesting object to study star formation and investigate the role of expanding \ion{H}{II} region in ongoing star-formation activity. The young stellar populations of the embedded cluster and the properties of the bubble/\ion{H}{II} region/cluster have not been studied in the literature.  In this work, we assume that the region is at the same distance (2.3 kpc) as the Carina Nebula \citep{walborn95}.
We present an analysis of this region using archival and published data from multiple wavelengths including near-IR from VCNS, mid-IR from {\it Spitzer} and  WISE, far-IR from ${\it Herschel}$ and radio-continuum data from  Molonglo Galactic Plane Survey (MGPS).  The rest of the paper is organized as:  \S{2} describes the archival data used in this work, \S{3} describes our results and \S{4} gives the conclusions of our work.   

\begin{figure}
\begin{center}
\includegraphics[width=7cm]{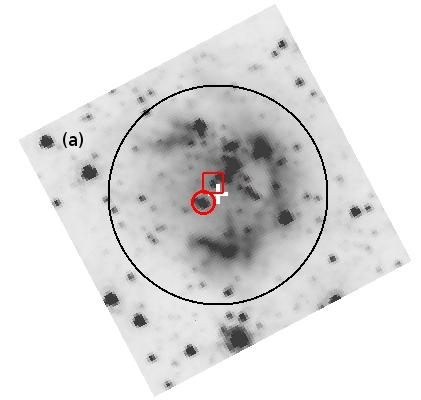}\par
\includegraphics[width=8cm]{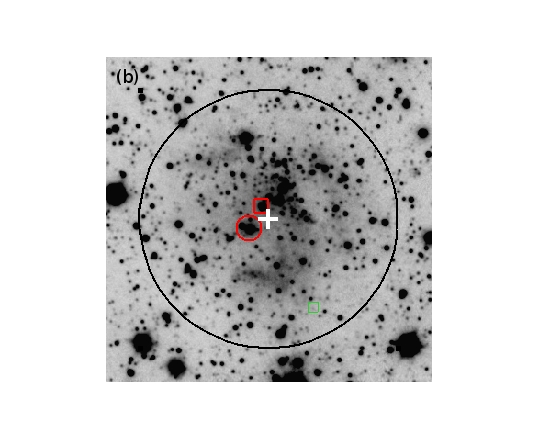}\par
\includegraphics[width=7cm]{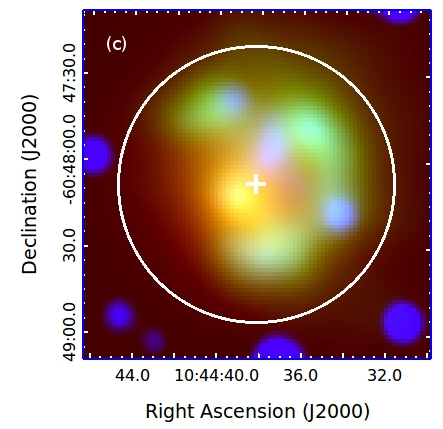}\par
\caption{The 2${\arcmin} \times$2${\arcmin}$ (a) {\it Spitzer} 4.5 $\mu$m image, (b) VISTA $K_s$-band image, and (c) the RGB (WISE 4.6 $\mu$m in blue, 12 $\mu$m in green, and 22 $\mu$m in red) image of the region.  Sources marked with box and circle on the {\it Spitzer} 4.5 $\mu$m and VISTA $K_s$-band image are discussed in the text (S 3.2).  The big circle of diameter 1\farcm6 ($\sim$ 1 pc assuming the distance to the region to be 2.3 kpc) shows the extent of the bubble.  The plus symbol marks the position of IRAS\,10427-6032.  The partial ring structure of the region is clearly visible in the {\it Spitzer} 4.5 $\mu$m and in WISE 12 $\mu$m. North is up and east is to the left in the image.}
\end{center}
\end{figure}

\section{Archival Data}
We study an area of 5${\arcmin} \times$5${\arcmin}$ around the position of the IRAS 10427-6032 for our analysis in this work.   
\subsection{Near-IR Data}
The VCNS survey \citep{preibisch14a} was conducted using the 4 m Visible and Infrared Survey Telescope for Astronomy \citep{emerson06} to obtain a deep 2 $\times$ 2 tile image mosaic covering a total sky area of $\sim$6.7 square-degrees ($\sim$ 2\fdg3 $\times$  2\fdg9) of the Carina Nebula, in the $J$, $H$, and $K_s$-bands. The final VCNS catalogue contains 3,951,580 sources detected in any two of the three bands with 5$\sigma$ magnitude limits for sources $\sim$20.0, 19.4, and 18.5 mag, in the $J$, $H$, and $K_s$ bands, respectively.  At the brighter end, stars with magnitudes less than  $J=$11.8, $H=$11.2, and $K_s=$10.5 mag,  are expected to be in the nonlinear or saturated regimes of the detectors. For these brighter stars, 2MASS magnitudes \citep{skrutskie06} are used.  The complete 5${\arcmin} \times$ 5${\arcmin}$ data were unavailable in the VCNS as the region lies near the edge of the observed VCNS field.  In particular, the survey did not cover the southern $\sim$1\farcm5 region.  We thus downloaded 5${\arcmin} \times$3\farcm5 catalog centered on the position of the IRAS\,10427-6032 from the \citep{preibisch14a} using the VizieR\footnote{http://vizier.u-strasbg.fr/viz-bin/VizieR} catalogue access tool.  We as well downloaded an image for the same field in the $K_s$ band from the VISTA archive\footnote{http://horus.roe.ac.uk/vsa/dbaccess.html}. For the southern 5${\arcmin} \times$1\farcm5 area that lacked coverage in the VCNS, we downloaded sources from the 2MASS catalogue.  The 2MASS Point Source Catalog \citep{skrutskie06} has the 10$\sigma$ detection limits of $J \sim$15.8 mag, $H \sim$ 15.1 mag, and $K_s \sim$ 14.3 mag.  

From the retrieved VCNS catalogue, there are 2570 detections in the $J$ band, 2715 in the $H$ band, and 2496 detections in the $K_s$ band.  For the purpose of our analysis of the near-IR color-color diagram and identification of YSO candidates, we discarded all detections in the $J$, $H$, and $K_s$ bands with signal-to-noise ratio (SN) $<$ 10.  This left us with 1888 sources in the $J$ band, 1958 in the $H$ band, and 1833 in the $K_s$ band from the VCNS.  From the 2MASS we downloaded 62 sources simultaneously detected in the three bands with SN $>$ 10. 

\begin{figure*}
 \begin{center}
  \includegraphics[width=\columnwidth]{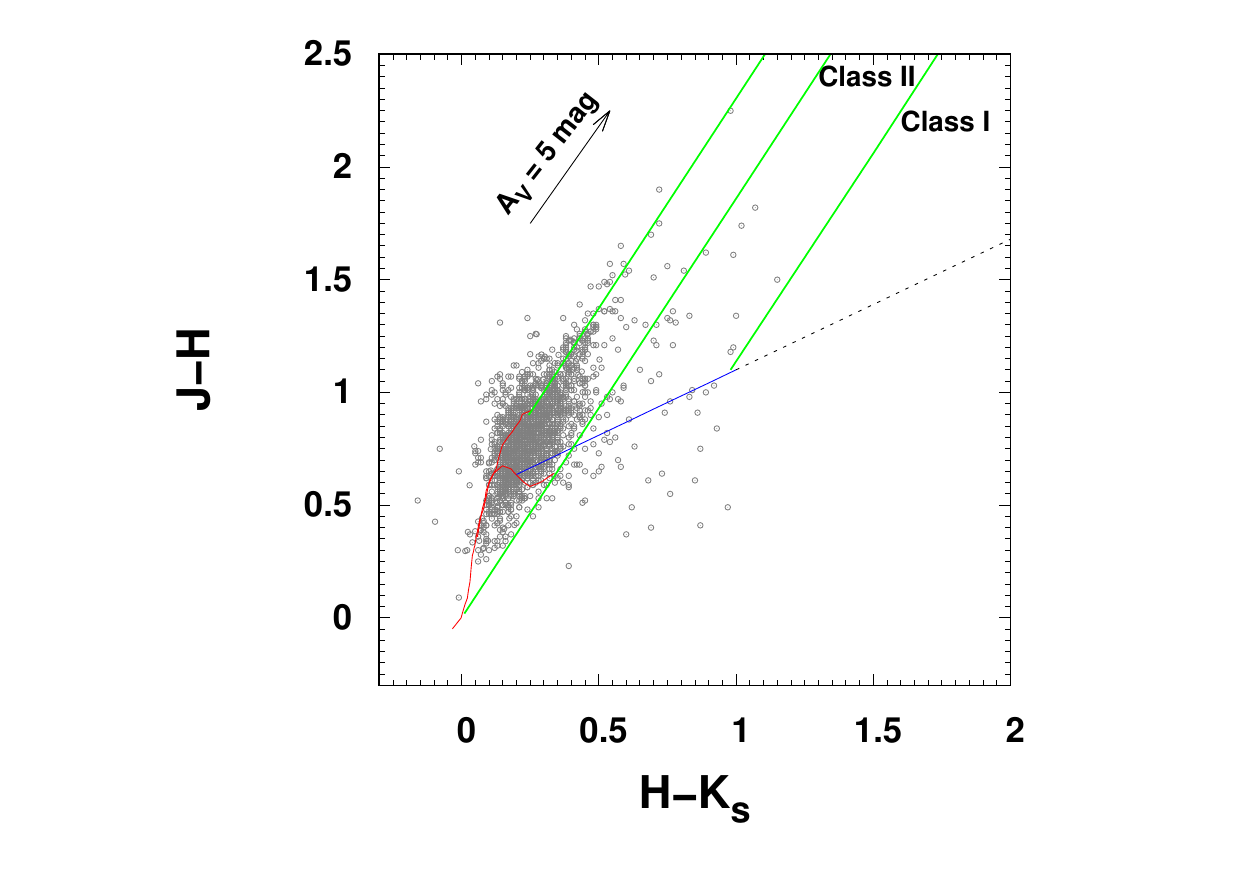}
    \includegraphics[width=\columnwidth]{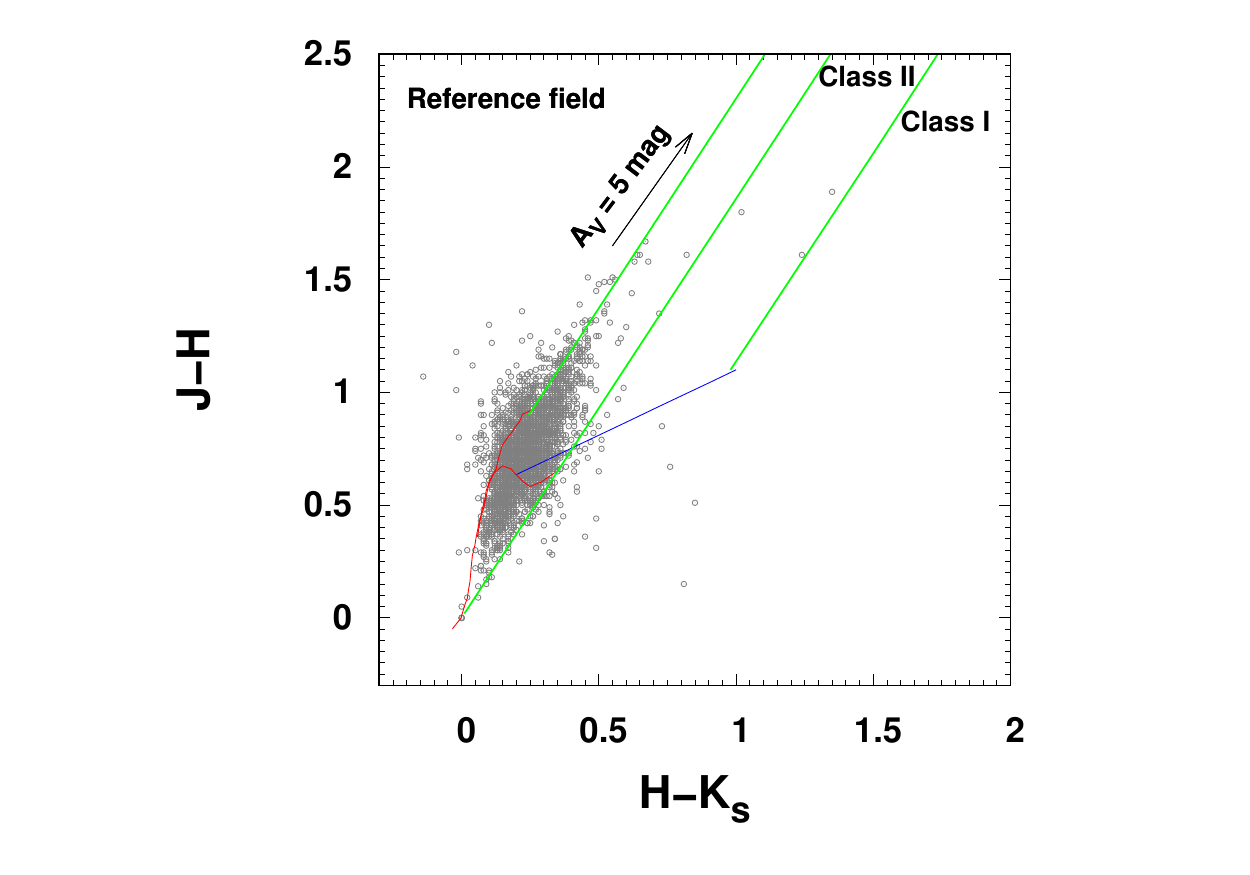}

  \caption{(a) The near-IR color-color diagram of sources detected in 5${\arcmin} \times$5${\arcmin}$ field centered on IRAS\,10427-6032 with SN $>$ 10 in VCNS and in 2MASS for a sub-region with no data within the VCNS.  Colors of main-sequence dwarfs and giants are from \citet{bessell88}, and the locus of unreddened CTTS is from \citet{meyer97}.  The reddening lines are plotted with a slope of 1.89 as determined by \citet{zeidler16} for this line-of-sight sources. The arrow depicts extinction A$_v$ = 5 Mag. The regions where Class I and Class II YSO candidates are found are marked on the graph. All sources are shown as grey open circles. (b) The near-IR color-color diagram of the reference field. All curves and lines are plotted in an identical manner as in (a).}
 \end{center}

\end{figure*}

\subsection{Mid-IR Data}
We made use of the data from two {\it Spitzer} surveys, the Vela Carina Survey \citep{majewski07}, and the Deep Glimpse Survey \citep{whitney11}.  The Vela Carina Survey covered the Galactic longitudes 255$^{\circ}$--295$^{\circ}$ for a latitude width of about 2$^{\circ}$, encompassing 86 square degrees of the Carina and Vela regions of the Galactic plane \citep{majewski07}. This area was observed in all the four IRAC bands centered at 3.6, 4.5, 5.8, and 8.0 $\mu$m.  For the Deep Glimpse project, {\it Spitzer} observed the regions, 25$^{\circ} <$ l $<$ 65$^{\circ}$, 0$^{\circ} <$ b $<$  +2\fdg7, and 265$^{\circ} <$ l $<$ 350$^{\circ}$, $-$2$^{\circ} <$ b $<$  +0\fdg1 in the two IRAC bands 3.6 and 4.5 $\mu$m only \citep{whitney11}. For both the surveys, two types of source lists are available for download in the InfraRed Science Archive\footnote{This research has made use of the NASA/ IPAC Infrared Science Archive, which is operated by the Jet Propulsion Laboratory, California Institute of Technology, under contract with the National Aeronautics and Space Administration.}, a highly reliable point source catalogue, and a more complete (but less reliable) point source catalogue.  The images available from the Vela Carina Survey of this region had IRAS 10427-6032 at the edge of the observed pointing (\#28850) in all the four bands.  Moreover, in the 4.5 $\mu$m and the 8.0 $\mu$m images, the observed field around our target had a defect and did not load upon downloading.  Due to these issues, we found no detections either in the highly reliable or in the most complete point-source catalogue in [4.5] and [8.0] bands. We thus largely made use of the 3.6 and 4.5 $\mu$m images and catalogues from the Deep Glimpse Survey in this work.

\subsection{Far-IR Data}
We used the  data taken with the  Photodetector Array Camera and Spectrometer \citep[PACS;][]{pogli10}
and  Spectral and Photometric Imaging Receiver \citep[SPIRE;][]{griffin10} of the {\it Herschel} Space Observatory, as a part of
the Proposal ID `OT1\,tpreibis\,1' (PI: Thomas Preibisch). For our analyses, we obtained the PACS 70 and 160 $\mu$m level-2.5  maps (processed with SPG v14.2.0) and the SPIRE 250, 350, and 500 $\mu$m extended calibrated level-3 (processed with SPG v14.1.0) maps for 5\farcm5 $\times$ 5\farcm5 area centered on IRAS 10427-6032, from the {\it {\it Herschel}} Science Archive\footnote{http://www.cosmos.esa.int/web/{\it Herschel}/science-archive}. The angular resolutions of these maps are  
10${\arcsec}$, 13${\arcsec}$, 20${\arcsec}$, 26${\arcsec}$ and 36${\arcsec}$ at 70, 160, 250, 350 and 500 $\mu$m, respectively \citep[see][]{pre12}. 
We note that the SPIRE maps are in the unit of  MJy sr$^{-1}$, while the PACS maps are in the unit of Jy pixel$^{-1}$.

\subsection{Radio Continuum Data}
The second epoch Molonglo Galactic Plane Survey (MGPS-2) carried out with the Molonglo Observatory Synthesis Telescope surveyed the Galactic longitudes 245$^{\circ}-$365$^{\circ}$ for Galactic latitudes $\lvert{b}\rvert<$ 10$^{\circ}$ ~at 843 MHz \citep{murphy07}.  The survey provides 4\fdg3 $\times$ 4\fdg3 mosaic images with 43$\times$43 cosec$\lvert{\delta}\rvert$ arcsec$^{2}$   resolution.  We downloaded the original processed image for this region from their website\footnote{http://www.astrop.physics.usyd.edu.au/mosaics/}.  We made use of the image to derive the physical parameters of the source, as well as to study the overall morphology of the region.

\section{Results}
\subsection{Identification of YSO candidates}
We made use of the near-IR and mid-IR data to identify YSO candidates of the region.  We first plotted a J$-$H vs. H$-K_s$ color-color diagram (see Figure~2a) of sources detected in all the three bands, $J$, $H$, and $K_s$ with SN $>$ 10. There are a total of 2030 such sources (1968 from VCNS and 62 from 2MASS). The reddening lines are plotted using a slightly steeper value of the slope of the reddening law, 1.86, as compared to 1.69 from \citet{rieke85}.  This value of the slope was found to most accurately fit the particular line of sight sources for the complete 6.7 square degree field encompassing the whole Carina Nebula by \citet{zeidler16}.  The reddening lines originating from the tip of the giant branch and the root of the main-sequence dwarf locus forms the main-sequence reddening band. Sources falling in this reddening band are likely field stars or evolved population of cluster members with little or no near-IR excess \citep{lada92}.  Those falling beyond this reddening band, on the redder side, are the ones exhibiting near-IR excesses.  We plotted a third reddening line originating from the tip of the empirical CTTS locus \citep{meyer97}.  Regions occupied by the reddened Class II and Class I YSOs  \citep{lada92} are labeled in the figure.  We found 68 sources with near-IR excess, of which 23 are Class II candidates.  For comparison, the near-IR color-color diagram of the same size reference field as the target field, centered on $RA=$161\fdg07337, $Dec.=-$60\fdg74447, shows only 3 sources in the region occupied by reddened Class II YSOs (Figure~2b). The remaining 45 sources with near-IR excesses are occupying a region where Herbig Ae/Be stars are found \citep{hillenbrand92}.  Some of these sources could be Herbig Ae/Be type stars, however, comparison with Fig.~2b suggests a fraction of them could as well be contaminants or evolved Ae/Be stars.  We thus do not include these 45 sources with small near-IR excesses in our discussion henceforth.  

\begin{figure}
 \begin{center}
  \includegraphics[width=\columnwidth]{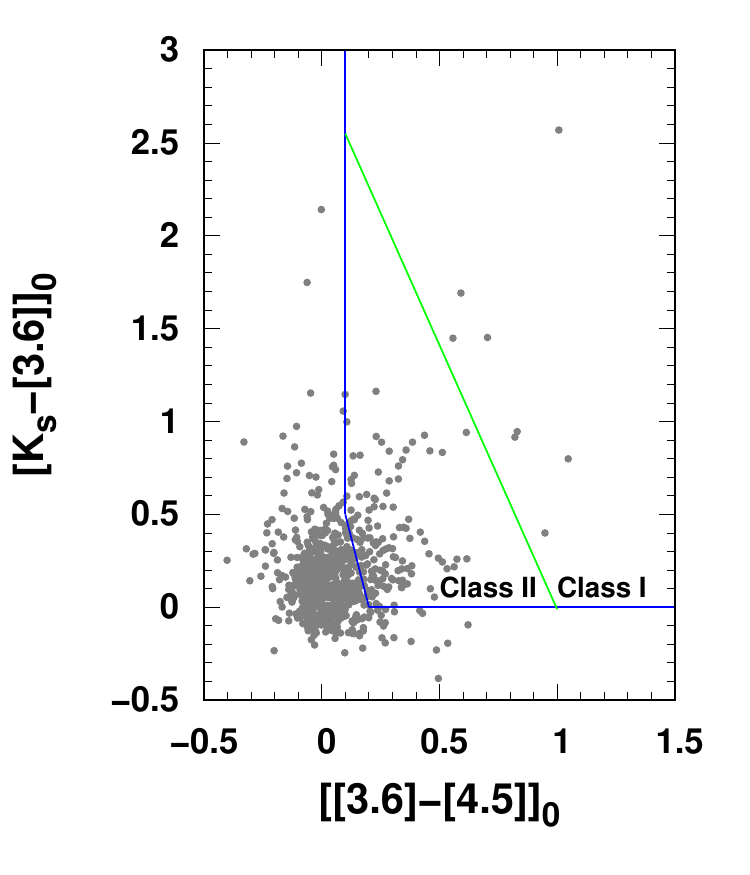}
  \caption{The color-color diagram of dereddened K$_s-$[3.6] versus [3.6]$-$[4.5] colors of all detections with SN $>$ 10 for the 5${\arcmin} \times$5${\arcmin}$ field centered on IRAS\,10427-6032.  The regions occupied by Class I and Class II YSO candidates according to the \citet{gutermuth09} criteria are demarcated on the figure.}
 \end{center}

\end{figure}

Our field of interest (5${\arcmin} \times$ 5${\arcmin}$ around the IRAS source) did not have a complete coverage in the Vela-Carina Survey, and additionally suffered a defect in the 4.5 and 8.0 $\mu$m images.  The Deep Glimpse survey, on the other hand, was carried out in only two of the four IRAC bands, 3.6 and 4.5 $\mu$m.  Thus, we could not make use of any color-color diagram involving mid-IR bands alone to identify YSO candidates \citep{allen04,gutermuth09}. We combined the mid-IR data from the Deep Glimpse Survey of {\it Spitzer} with the near-IR data from the VCNS to identify additional YSO candidates.  We cross-matched the sources from the two catalogues within 1${\arcsec}$ search radius to identify counterparts.  The color-color diagram of unreddened K$_s-$[3.6] vs. [3.6]$-$[4.5] colors of the cross-identified sources is shown in Figure~3. In order to estimate the unreddened $K_s-$[3.6] and [3.6]$-$[4.5] colors from the corresponding observed colors, we first used the J$-$H vs. H$-K_s$ color-color diagram (Figure~2a) to find the line-of-sight extinction to the region.  In particular, we dereddened all the sources falling in the reddening band up to a baseline, plotted parallel to the main-sequence dwarf locus (K5--M5), to find the color excesses, E(H$-K_s$).  Then we estimated the dereddened K$_s-$[3.6] colors of sources using the relation, ${E_{(H-K_s)}}/{E_{(K_s-[3.6])}}=1.49$, and the dereddened [3.6]$-$[4.5] colors of sources using the relation, ${E_{(H-K_s)}}/{E_{([3.6]-[4.5])}}=1.17$ \citep{flaherty07}. To select Class II YSO candidates from this plot, we followed \citet{gutermuth09} criteria as detailed below:
{\footnotesize 
\begin{align*}
[[3.6]-[4.5]]_0 - \sigma_1  & >  0.101 \\
[K_s-[3.6]]_0 - \sigma_2 & >  0  \\
[K_s-[3.6]]_0 - \sigma_2 & >  -2.85714 \times ([[3.6]-[4.5]]_0 - \sigma_1 - 0.101) + 0.5 
\end{align*}
}%
Here $\sigma_1$ and $\sigma_2$ are found using photometric errors in the three magnitudes $K_s$, [3.6], and [4.5], by error propagation as,
$\sigma_1=\sigma{[[3.6]-[4.5]};\sigma_2=\sigma{[[K_s]-[3.6]]}$.
There are 40 sources that satisfied this set of criteria so are Class II YSO candidates.  Five of these sources are found to satisfy an additional criterion and are Class I YSO candidates:
{\footnotesize 
\begin{align*}
[K-[3.6]]_0 - \sigma_2 > -2.85714 \times ([[3.6]-[4.5]]_0 - \sigma_1 -0.101) + 0.5
\end{align*}
}%
To further reduce dim extragalactic contaminants, we imposed a cut based on the unreddened 3.6 $\mu$m mag as employed by \citet{gutermuth09}.  We discarded all sources with [3.6]$_{0} > 15$ mag for Class II candidates and [3.6]$_{0} > 14.5$ mag for Class I candidates.  This leaves us with a total of 11 candidates of which 5 are Class I and 6 are Class II candidates.  Combining with YSO candidates from Figure~2a, we have a total of 29 Class II and 5 Class I candidates. Our strict criteria may eliminate some of the genuine YSOs of the region, however we prefer to use this secure sample of YSO candidates to study the region. 

\citet{marton16} employed support vector machine algorithm to determine YSO candidates using all-sky data from WISE and 2MASS.  Seven of their YSO candidates are found in the region studied in this work. In our classification scheme, we retrieved 3 of these YSO candidates, whereas the remaining four sources turned out to be non-YSOs according to our criteria. We made use of the multi-wavelength information to ascertain the nature of these four sources.  One of these sources ($RA$ = 161\fdg1763118, $Dec.=-$60\fdg792419) lacked detection in the {\it Spitzer} IRAC [3.6] and [4.5] bands and thus could not be a YSO.  Two sources $RA$ = 161\fdg390312, $Dec.=-$60\fdg788896, and $RA$ = 161\fdg1298658, $Dec.=-$60\fdg8153006, did neither show excess in near- and mid-IR, nor suffered reddening.  These sources were found near the main-sequence branch on the near-IR color-color diagram.  These sources are thus also ruled out as YSO candidates.  The fourth source, $RA$ = 161\fdg1311654, $Dec.=-$60\fdg810078, was found in the main-sequence reddening band in the near-IR color-color diagram, however did not show excess in the mid-IR, 3.6 and 4.5 $\mu$m.  Though it could be an evolved YSO such as Class III, since it does not fit in our YSO identification criteria, to be consistent we do not consider it a YSO candidate.  We thus conclude that only three of the seven  YSO candidates from \citet{marton16} of this region are likely YSOs. Our final list of YSO candidates thus contain 29 Class II candidates and 5 Class I candidates.  The ratio of Class II/Class I candidates ($\sim$ 6) indicate that this is a young star-forming region.    

 
\begin{table*}
	\caption{Main parameters from SED analysis}
	\centering
	{\footnotesize
	\begin{tabular}{llllllllllll}
		\hline
		\\
	    No.&Class&RA (deg) &Dec. (deg) &log $t_{*}$&Mass&log $M_{disk}$&log $\dot{M}_{disk}$&log $T_{*}$
		&log $L_{total}$&$A_{V}$&$\chi^{2}_{min}$\\
		 &~ &(J2000)&(J2000)&(log yr)&($M_{\odot}$)&(log $M_{\odot}$)&(log $M_{\odot}~yr^{-1}$)&(log K)&(log $L_{\odot})$&(mag)&(per data point)\\
		 \\
		\hline
		\\
1&I&161.142605 & -60.804943 & 4.09$\pm{0.64}$ & 3.01$\pm{1.74}$ & $-$1.88$\pm{0.80}$ & $-$6.40$\pm{1.11}$ & 3.63$\pm{0.05}$ & 1.94 $\pm{0.42}$ & 16.82$\pm{3.4}$ & 4.84 \\
2&II&161.214739& -60.814748 & 5.48$\pm{0.78}$ & 2.75$\pm{1.55}$ & $-$1.52$\pm{1.32}$ & $-$7.73$\pm{1.48}$ & 3.74$\pm{0.89}$ & 1.57 $\pm{0.57}$ & 7.31$\pm{4.28}$ & 2.14 \\ 
3&II&161.163985& -60.793807 & 4.75$\pm{0.38}$ & 2.83$\pm{1.77}$ & $-$2.04$\pm{0.78}$ & $-$6.70$\pm{0.88}$ & 3.63$\pm{0.02}$ & 1.73 $\pm{0.41}$ & 10.9$\pm{0.42}$ & 5.57 \\      
4&II&161.163399& -60.794834 & 3.89$\pm{0.21}$ & 5.17$\pm{1.02}$ & $-$2.03$\pm{0.85}$ & $-$7.49$\pm{0.76}$ & 3.89$\pm{0.21}$ &2.50 $\pm{0.39}$ & 9.87$\pm{4.27}$ & 18.9 \\      
5&II&161.156170& -60.799027 & 6.58$\pm{0.26}$ & 3.89$\pm{0.62}$ & $-$2.45$\pm{0.71}$ & $-$8.36$\pm{1.12}$ & 4.14$\pm{0.04}$ & 2.26 $\pm{0.27}$ & 7.88$\pm{2.44}$ & 9.67 \\      
6&II&161.133327& -60.762207 & 6.35$\pm{0.40}$ & 2.91$\pm{0.69}$ & $-$4.37$\pm{1.17}$ & $-$10.1$\pm{2.10}$ & 3.86$\pm{0.18}$ & 1.51 $\pm{0.51}$ & 3.00$\pm{1.25}$ & 6.02 \\    
7&II/III&161.157004& -60.800308 & 5.48$\pm{0.78}$ & 2.75$\pm{1.55}$ & $-$1.52$\pm{1.32}$ & $-$7.73$\pm{1.48}$ & 3.74$\pm{0.89}$ & 1.57 $\pm{0.57}$ & 7.31$\pm{4.28}$ & 2.14 \\      
\hline
\end{tabular}\label{table:A01}
}%
\end{table*}

\begin{figure*}
 \begin{center}
  \includegraphics[scale=0.19]{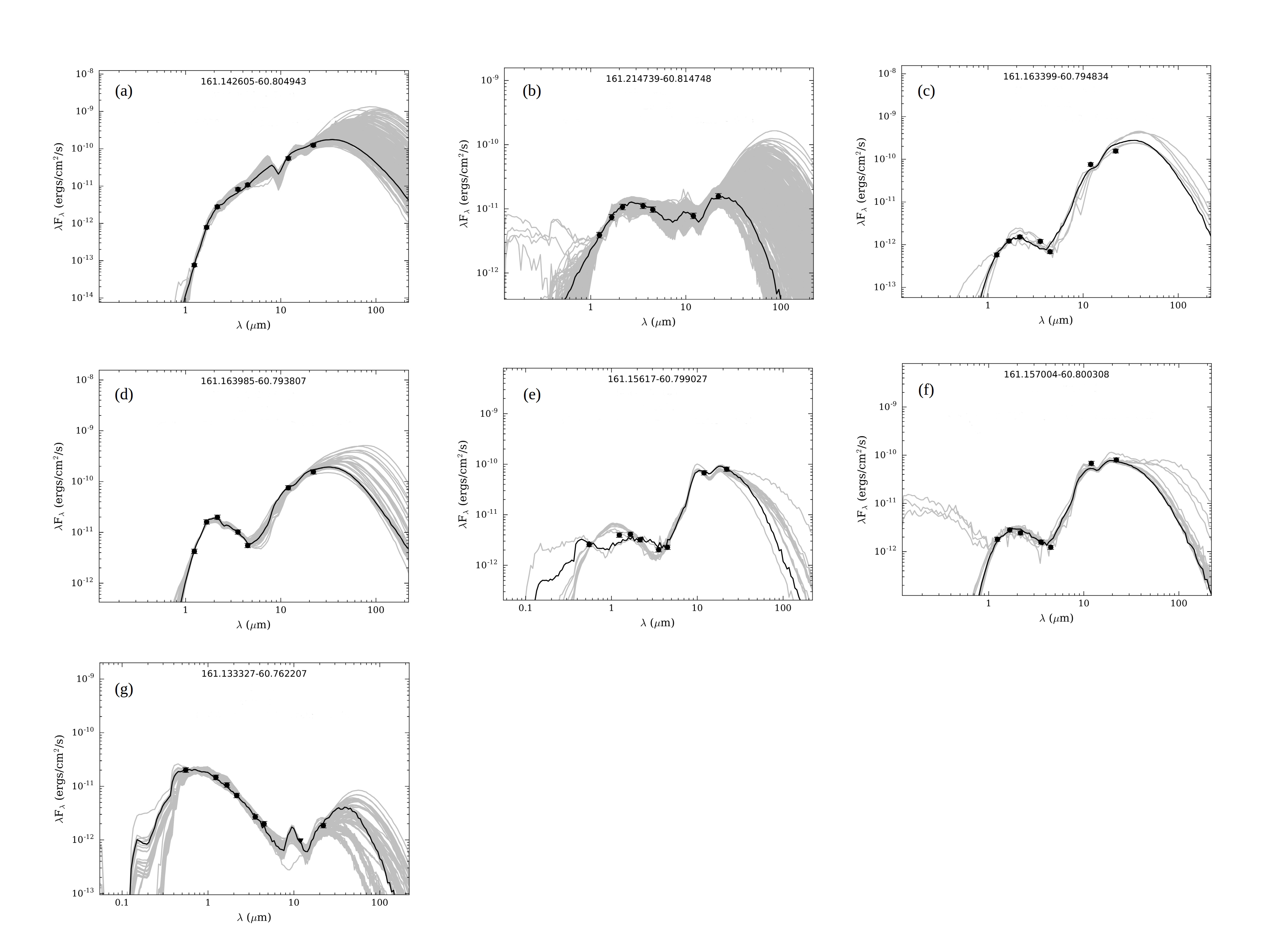}
  \caption{The best fit SEDs of YSO candidates (a) 161.142605-60.804943 (b) 161.214739-60.814748 (c) 161.163399-60.794834 (d) 161.163985-60.793807 (e) 161.15617-60.799027 (f) 161.157004-60.800308 and (g)  161.133327-60.762207.  The black dots mark the data points. The solid black curve is the best fitted model, while the grey curves denote the subsequent good fits for ${\chi}^2 -{\chi}_{min}^{2}$ (per data point) $<$ 3.}
 \end{center}
\end{figure*}

\subsubsection{Mass Distribution and Spectral Energy Distribution (SED) Fitting of Selected YSO candidates}

To characterize and understand the nature of YSOs in the cluster, we constructed the SEDs of YSO candidates for which photometric magnitudes are available in seven or more bands.  There are 7 such sources.   We fitted SEDs to our candidates using the grid of models and fitting tools as described by \citet{robitallie06} using their \textit{SED Fitter python package}\footnote{https://github.com/astrofrog/sedfitter}. These models were computed using Monte Carlo based 20000 2D radiation transfer models from \citet{whitney03a} and by adopting several combinations of the central star, disc, infalling envelope, and bipolar cavity, for a reasonably large parameter space. Their total YSO grids consist of 200,000 SEDs as each of the 20000 YSO models have SEDs for ten different inclination angles. This tool provides various physical parameters of the YSOs making it an ideal tool to study the evolutionary status of YSOs  in star-forming regions.

We used the photometric magnitudes of the YSO candidates in the $J$, $H$, $K_s$ from VCNS, 3.6 and 4.5 $\mu$m from {\it Spitzer} Deep Glimpse Survey, and 12 and 22 $\mu$m from WISE W3 and W4 filters.  Out of 7, for two of our YSO candidates, we could unambiguously find an optical counterpart in the DSS red image.  For these sources,  we thus had the 8th data point, namely the $V$ mag of the source, for the SED fitting. The WISE filters W1--W4 are not available in the \textit{SED Fitter python package}.  We thus prepared WISE W3 and W4 filters as prescribed in \textit{SED Fitter python package} using a class `Filter' to perform broadband convolution, to obtain the convolved fluxes.  To do so we used the per-photon relative system response curves of W3 and W4 bands from \citet{wright10}. While fitting the SEDs we set photometric uncertainties to be 10\% of the magnitudes instead of the formal photometric errors in order to fit without any possible bias caused by an underestimate in the flux uncertainties. Since the distance estimates of the clusters in the Carina Nebula have large uncertainties, 2.0--4.0 kpc \citep{bakkar80,carraro04,degi01,hur12}, partly due to abnormal reddening law \citep{feinstein73,carraro04}, we used a range of 2.0--3.5 kpc as our input to fit the SEDs.  For the extinction, we used a range of 0--18 mag as the maximum extinction suffered by the sources in our region of study was found to be $A_V \sim$ 18 mag. Finally we considered models with $\chi^2-\chi^2_{min}$ (per data point) $<$ 3 relative to the model of best-fit for our analysis.  The physical parameters from the best fitted models are given in Table~1.  The resultant SED fits to the YSO candidates are given in Figure~4.  The masses from the SED fitting of the YSO candidates vary from 2.7 to 5.1 M$_{\odot}$.  The median age of all YSO candidates is $\sim$0.17 Myr.  Two YSO candidates, \#1 and \#4 in Table~1 are the youngest sources with estimated ages $\sim$10$^{4}$ years.  Based on the shape of the SED, the YSO candidate \#1 in Table~1 ($RA$=161\fdg142605, $Dec.$=$-$60\fdg804943) is a Class I YSO which is also consistent with its identification based on both near- and mid-IR color excess criteria as defined in \S{3.1}.  The only other Class I YSO candidate for which the SED was constructed is \#2 in Table~1 ($RA$=161\fdg214739, $Dec.$=$-$60\fdg814748).  The SED for this candidate shows a flat spectrum with an estimated age from the SED fitting, $\sim$0.3 Myr.   This YSO thus appears to be somewhat more evolved compared to \#1.  All the other YSO candidates whose SEDs are presented were identified to be Class II YSO candidates and show Class II type SEDs.   

\begin{figure}
  \includegraphics[width=1.1\columnwidth]{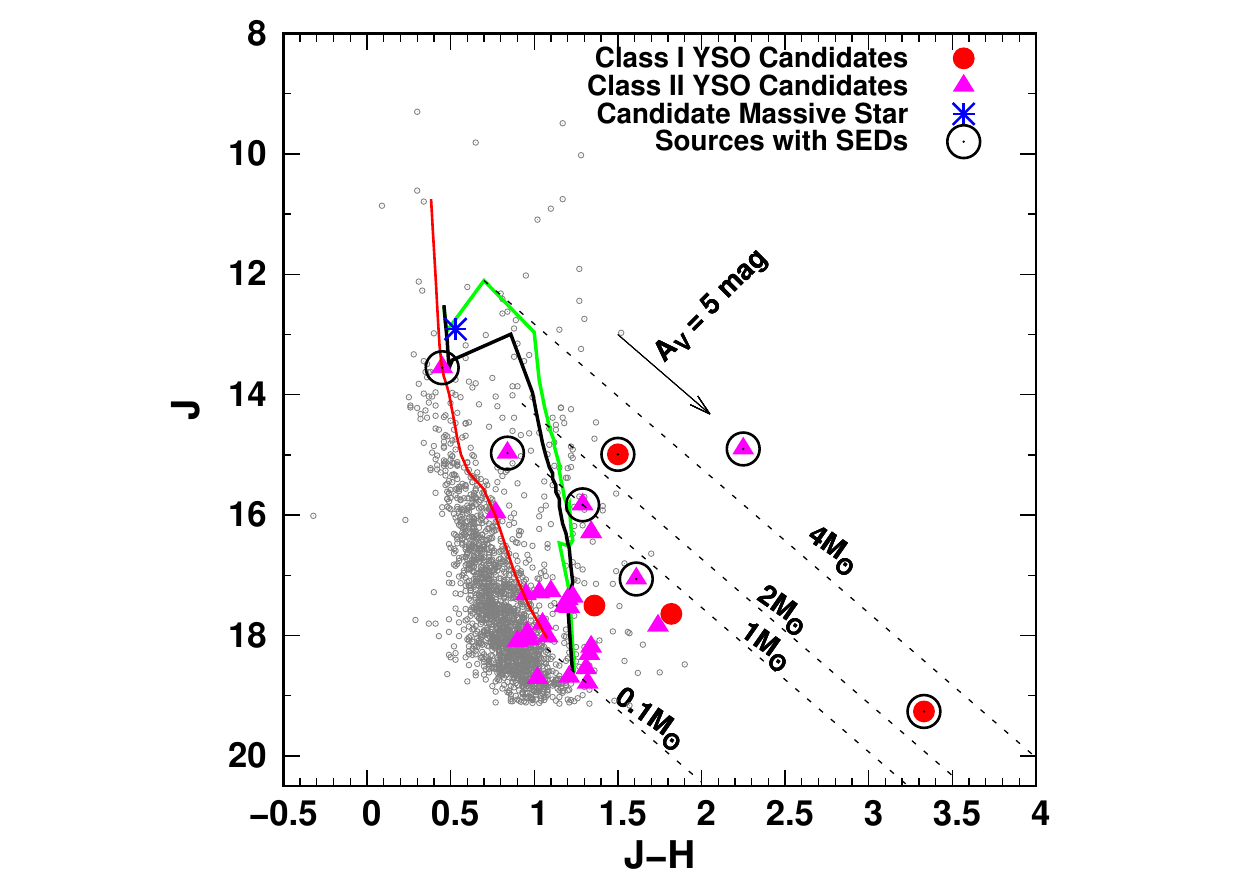}
  \caption{The color-magnitude diagram $J$ vs. $J-H$ of all sources (grey open circles) detected in 5${\arcmin} \times$5${\arcmin}$ field centered on IRAS\,10427-6032 with SN $>$ 10 in VCNS and in 2MASS for a sub-region with no data within the VCNS.  The Class II YSO candidates are shown as filled triangles, Class I YSO candidates are shown as filled circles and the candidate massive star is shown as a blue asterisk. Sources for which SEDs are constructed are marked with large open circles.  The most massive members of the region are fitted with a 1 Myr main-sequence isochrone (red color) of the Geneva stellar tracks \citep{lejeune01} after dereddening it by $A_V$ = 5.0 mag, whereas low-mass YSO candidates are fitted by pre-main sequence isochrones of 1 Myr (green color) and 2 Myr (black color) from \citet{siess00} for the mass-range of 0.1 M$_{\odot}$ to 7.0 M$_{\odot}$. One of the Class I YSO candidates has no detection in the $J$ band and thus is not plotted on the graph.}
\end{figure}

\begin{figure*}
\begin{center}
\includegraphics[width=\columnwidth]{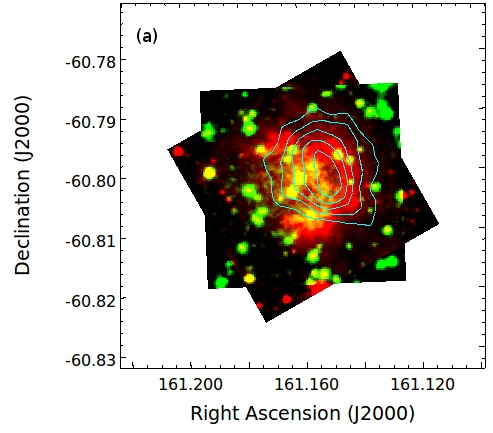} 
\includegraphics[width=\columnwidth]{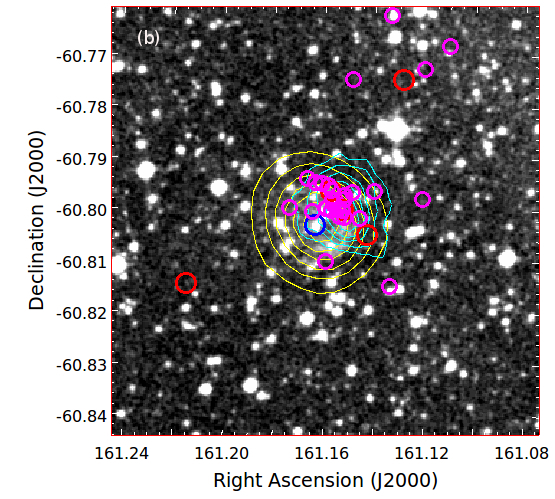} 
\caption{(a) The two-color 2${\arcmin} \times$2${\arcmin}$ field around IRAS\,10427-6032 with the DSS-2 $r$-band in green and the $\mathit{Spitzer}$ IRAC 4.5 $\mu$m in red.  The 870 $\mu$m contours from ATLASGAL are overlaid on the image. (b) The 5${\arcmin} \times$5${\arcmin}$ field around IRAS\,10427-6032 in the DSS-2 r-band. The 843 MHz contours from the MGPS-2 (yellow) and the 870 $\mu$m contours from the ATLASGAL (cyan) are overlaid on the image. 
The YSO candidates are marked as magenta circles (Class II) and red circles (Class I).  The candidate massive star is marked as a blue circle.}
\end{center}
\end{figure*}

Figure~5 shows the near-IR color-magnitude diagram, J vs. J$-$H, of all near-IR sources in our sample. A 1-Myr main-sequence isochrone of the Geneva stellar tracks \citep{lejeune01} is overplotted after reddening it by $A_V$ = 5 mag and assuming the distance to be 2.3 kpc.  The value of extinction was determined using the photometry of the candidate massive star. As discussed in \S{3.2}, the flux of ionizing photons responsible for the \ion{H}{II} region suggests the expected spectral type of the star to be $\sim$B0--B0.5.  We adopt the colors of the late-O and early-B main-sequence stars to determine color excesses, $E(J-H)$ and $E(H-K)$, of the candidate massive star. That gave us a range, $A_V$ = 4.5--5.5 mag, for the extinction suffered by the candidate massive star. We thus used the median value, $A_V$ = 5 mag, to fit the 1-Myr main-sequence isochrone. Our low-mass YSO candidates are found to cluster near a 1-2 Myr PMS isochrones \citep{siess00} drawn for the mass-range of 0.1 M$_{\odot}$ to 7.0 M$_{\odot}$.  YSO candidates for which SEDs are constructed show general agreement in parameters derived based on SED fitting and isochrones.  Two of the most evolved YSOs based on the age estimation from SED (\#5 and \#6 in Table~1) are appearing close to or on the main-sequence.  The mass estimates based on PMS isochrones for roughly half of the YSO candidates are consistent with the estimates derived from the SED fitting (2--5 M$_{\odot}$) whereas show some deviation for the remaining candidates.

\subsubsection{Spatial Distribution of YSO Candidates}
Figure~6a shows a 2$\arcmin \times$2$\arcmin$ two-color image centered on IRAS 10427-6032 with Digitized Sky Survey-2 (DSS-2) r-band in green and ${\mathrm Spitzer}$ IRAC [4.5] in red.  The cold dense clump detected at 870 $\mu$m (shown as contours) is seen to be adjacent to the optical nebula.  Figure~6b shows both the 843 MHz radio continuum emission as well as the 870 $\mu$m emission contours overlaid on a 5$\arcmin \times$5$\arcmin$ field centered on the IRAS object on the DSS-2 r-band image.  As can be seen, the cold dust clump is protruding into the ionized region.  Most of the YSO candidates are spatially coincident with the sub-mm contours.  A small number of the remaining YSO candidates are found to lie on the north-western side of the bubble rim whereas a single Class I YSO candidate is found on the eastern side of the bubble.  Out of the five Class I YSO candidates, two are found on each side of the bubble, whereas three are coincident with the bubble rim and the cold dust condensation.  

\begin{figure}
 \begin{center}
  \includegraphics[width=\columnwidth]{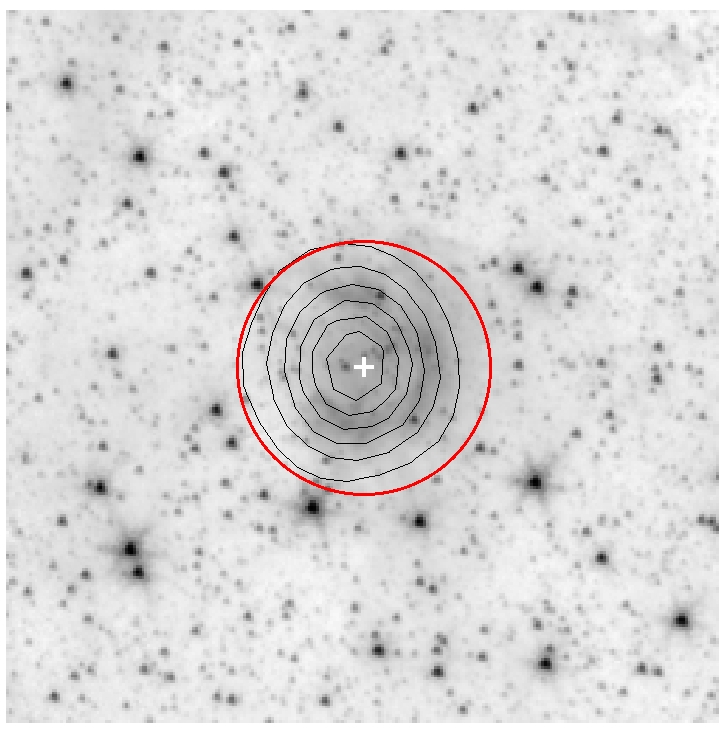}
  \caption{The {\it Spitzer} 5${\arcmin} \times$5${\arcmin}$ 3.6 $\mu$m image centered on IRAS\,10427-6032.  The contours of 843 MHz radio continuum data from MGPS-2 are overlaid on the image.  The position of the IRAS source is marked with a plus symbol.  The big circle shows the extent of the bubble and has a diameter of 1\farcm6.}
 \end{center}
\end{figure}

\subsection{Physical Properties of Compact H~II Region and Massive Star Candidate}

Figure~7 shows the {\it Spitzer} 3.6 $\mu$m image with 843 MHz MGPS-2 contours over-plotted on it.  The ionized emission shows a nearly spherical morphology which fills the bubble interior almost completely.  We used the AIPS tasks JMFIT, MAXFIT and IMEAN, on the 843 MHz image to fit the compact core of the ionized emission with a Gaussian model.  The obtained results are presented in Table~2.  The angular extent of the nearly spherical source is found to be $\sim$110$\arcsec$, which translates to a linear diameter of 1 pc assuming the cluster to be located at 2.3 kpc.  We determined the Lyman continuum luminosity (in photons ${\mathrm s^{-1}}$ ) required to generate the observed flux density using \citet{kurtz94} formula:

\[S_*  \geqslant \bigg(\frac {7.59 \times 10^{48}}{a(\nu,T_e)}\bigg)\bigg(\frac {S_{\nu}}{\mathrm Jy}\bigg)\bigg(\frac{T_e}{\mathrm K}\bigg)^{-0.5}\bigg(\frac{D}{\mathrm kpc}\bigg)^{2}\bigg(\frac {\nu}{\mathrm GHz}\bigg)^{0.1}\]
where $S_{\nu}$ is the integrated flux density in Jy, D is the distance in kiloparsec, $T_e$ is the electron temperature, a($\nu$, $T_e$) is the correction factor, and $\nu$ is the frequency in GHz at which the luminosity is to be calculated. We assumed $T_e$ to be the typical 10000 K, implying $a(\nu, T_e)$=0.99  as seen from Table 6 of \citet{mezger67}. The Lyman continuum luminosity is found to be $S_*$=$10^{46.9}$ photons $\mathrm{s^{-1}}$. 
To estimate the dynamical age of the \ion{H}{II} region ($t$), we used the following formula from \citet{spitzer78}
\[R(t) = R_s\bigg(1 + \frac{7c_{II}t}{4R_s}\bigg)^{\frac{4}{7}}\]
where $R(t)$ is the radius of the \ion{H}{II} region at time $t$, c$_{II}$ is the speed of sound in the \ion{H}{II} region taken to be 11 $\times 10^5$ cm $s^{-1}$ from \citet{palla05} and $R_s$ is the Stromgren radius \citep{stromgren79}, given by, 
\[R_s = \bigg(\frac {3S_*}{4\pi n_o ^2 \beta_2}\bigg)^\frac{1}{3}\]
In the above expression, $n_o$ is the initial ambient density in $\mathrm{cm^{-3}}$, and $\beta_2$ is the total recombination coefficient to the first excitetd state of hydrogen. We assumed $\beta_2$ to be $2.6 \times 10^{-13} \mathrm{cm^{3}~s^{-1}}$ \citep{palla05}. To estimate $n_0$, we used the gaseous mass of the \ion{H}{II} region (see \S{3.3}), and the measured size of the \ion{H}{II} region.  By assuming a uniform density throughout the \ion{H}{II} region, we deduced $n_o =$ 9.3 $\times$ 10$^{3}$ cm$^{-3}$.  However, this value of $n_0$ must be only treated as a lower limit of the actual density since some of the gaseous mass has already converted into stars and some of it has been ionized by the \ion{H}{II} region.  The dynamical age of the 
\ion{H}{II} region using this value of $n_0$ turns out to be $t$=0.30 Myr.  If we use an upper limit for the density instead, say $n_o=$ 10$^{5}$ cm$^{-3}$, the dynamical age of the \ion{H}{II} region turns out to be $t$=0.95 Myr.
\citet{tremblin14} studied the evolution of \ion{H}{II} regions in turbulent molecular clouds.  We estimated the dynamical age of the \ion{H}{II} region also using the method outlined in \citet{tremblin14}.  Comparing the parameters of IRAS\,10427-6032 with the pressure-size tracks in \citet{tremblin14}, the age of the \ion{H}{II} region is $\sim$0.5 Myr.  However, we note that this age is a lower limit as the method of \citet{tremblin14} is more appropriate for classical \ion{H}{II} regions where effects of magnetic fields and gravity are less important.  Based on both the methods, thus, the range of dynamical age of the \ion{H}{II} region is 0.3--0.95 Myr.  
A comparison of $\mathrm{log} S_*$ value with the values from \citet{panagia73}, assuming a ZAMS, suggests a spectral type of B0.5--B1 for the ionizing source.  

\begin{table}
\begin{center}
\caption{Fitting Results of Compact \ion{H}{II} Region}
\begin{tabular}{ll}
\hline					    
2D Gaussian fit size  		  & $56.62'' \times 53.30''$  \\ 
Position angle (deg)              & 67.446  \\ 
Peak flux density (mJy beam$^{-1}$)  & 133.93   \\ 
Integrated flux density (mJy)      & 171.5    \\ 
\hline
\end{tabular}
\end{center}
\end{table} 

There are two bright sources nearby the position of IRAS 10427-6032, a source $RA$=161\fdg16058, $Dec.$=$-$60\fdg80089 at an angular distance of $\sim$5$^{\arcsec}$, and another source $RA$=161\fdg16329, $Dec.$=$-$60\fdg80317 at an angular distance of $\sim$9$^{\arcsec}$.  These bright sources are marked with a box and a circle, respectively in Figure~1. The closer source ($RA=$161\fdg16058, $Dec.=-$60\fdg80089) is detected only in IRAC 3.6 $\mu$m and 4.5 $\mu$m, and becomes too faint to be visible in the 5.8 $\mu$m image from the Vela-Carina Survey.  It is also too faint to be visible in the WISE 12 and 22 $\mu$m images.  The farther source ($RA=$161\fdg16329, $Dec.=-$60\fdg80317) appears to be a candidate massive star responsible for the \ion{H}{II} region as it is seen to brighten up from the $K_s$ band to {\it Spitzer} IRAC 3.6 $\mu$m and 4.5 $\mu$m to WISE 12 $\mu$m and remains the only visible bright source in WISE 22 $\mu$m in the studied region.  We constructed the SED of this source using its photometric magnitudes/fluxes in eight bands, optical $B$, $V$, and $I$, VISTA $J$, $H$, $K_s$, and {\it Spitzer} IRAC 3.6 and 4.5 $\mu$m.  The fitted SED is shown in Figure~8. The best fit model suggests a spectral type B0--B0.5 of the source and an effective temperature of 25,000 K.  This is consistent with the ionizing flux of the \ion{H}{II} region.   

\begin{figure}
 \begin{center}
  \includegraphics[width=\columnwidth]{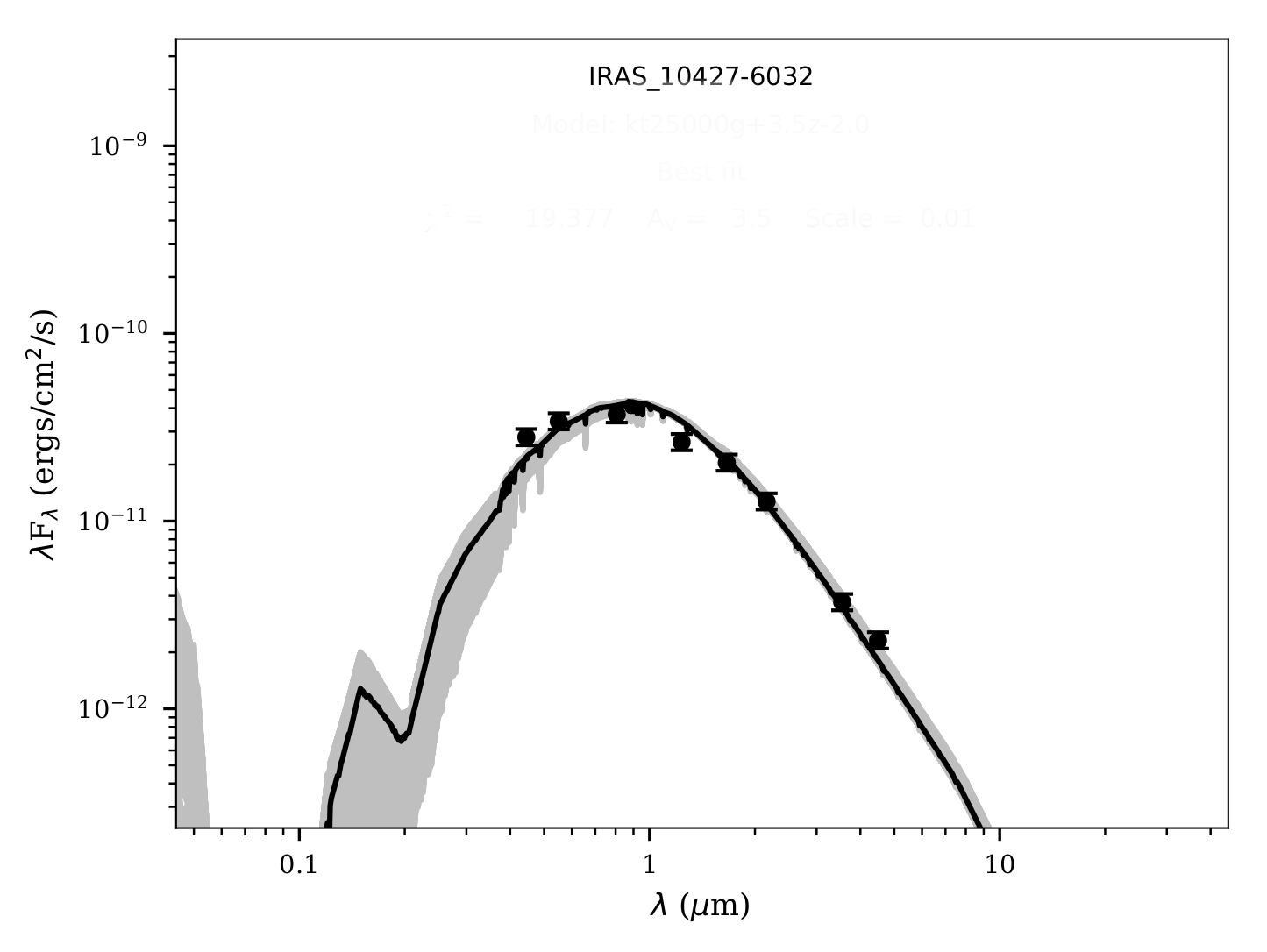}
  \caption{The SED fitting of the zero-age main-sequence candidate massive star.  The best-fit model suggests an effective temperature of 25,000 K. The black dots mark the data points. The solid black curve is the best fitted model, while the grey curves denote the subsequent good fits for ${\chi}^2 -{\chi}_{min}^{2}$ (per data point) $<$ 3. }
 \end{center}
\end{figure}

\subsection{{\it Herschel} Column Density and Temperature Maps}

{\it {\it Herschel}} observations with a wide wavelength coverage can reveal the dust properties  of a cloud complex. To 
explore dust properties around IRAS 10427-6032, we derived the column density and temperature maps by performing 
a pixel-to-pixel modified black-body fit to the 160, 250,
350 and 500 $\mu$m {\it {\it Herschel}} images following the procedure outlined in \citet{mal15}. 
Prior to performing the modified black-body fit, we first converted  all the SPIRE images to the PACS flux unit (i.e., Jy pixel$^{-1}$). 
Then, we convolved and regrided all the shorter wavelength images to the resolution and pixel size of the 500 $\mu$m map.
Next, to minimize the contribution of possible excess 
dust emission along the line-of-sight, we subtracted the corresponding background flux, estimated from a field nearly devoid of emission, from each image.   
In the final step, we fitted the modified black-body on 
these background subtracted images using the formula given in \citet{mal15}. While fitting, 
we used a dust spectral index of $\beta$=2, and the dust opacity per unit mass column density ($\kappa_{\nu}$) as given in 
\citet{beck91}, $\kappa_{\nu} = 0.1~(\nu/1000~{\rm GHz})^{\beta}$ cm$^2$/gm, leaving the dust temperature 
(T$_{dust}$) and dust column density (N(H$_{2})$) as free parameters. Since we are more interested in cold dust properties, to avoid contribution from stochastically heated small grains \citep[]{comp10,pav13}, we did not use 70 $\mu$m data in the fitting procedure.

\begin{figure}
\centering{
\includegraphics[width=\columnwidth]{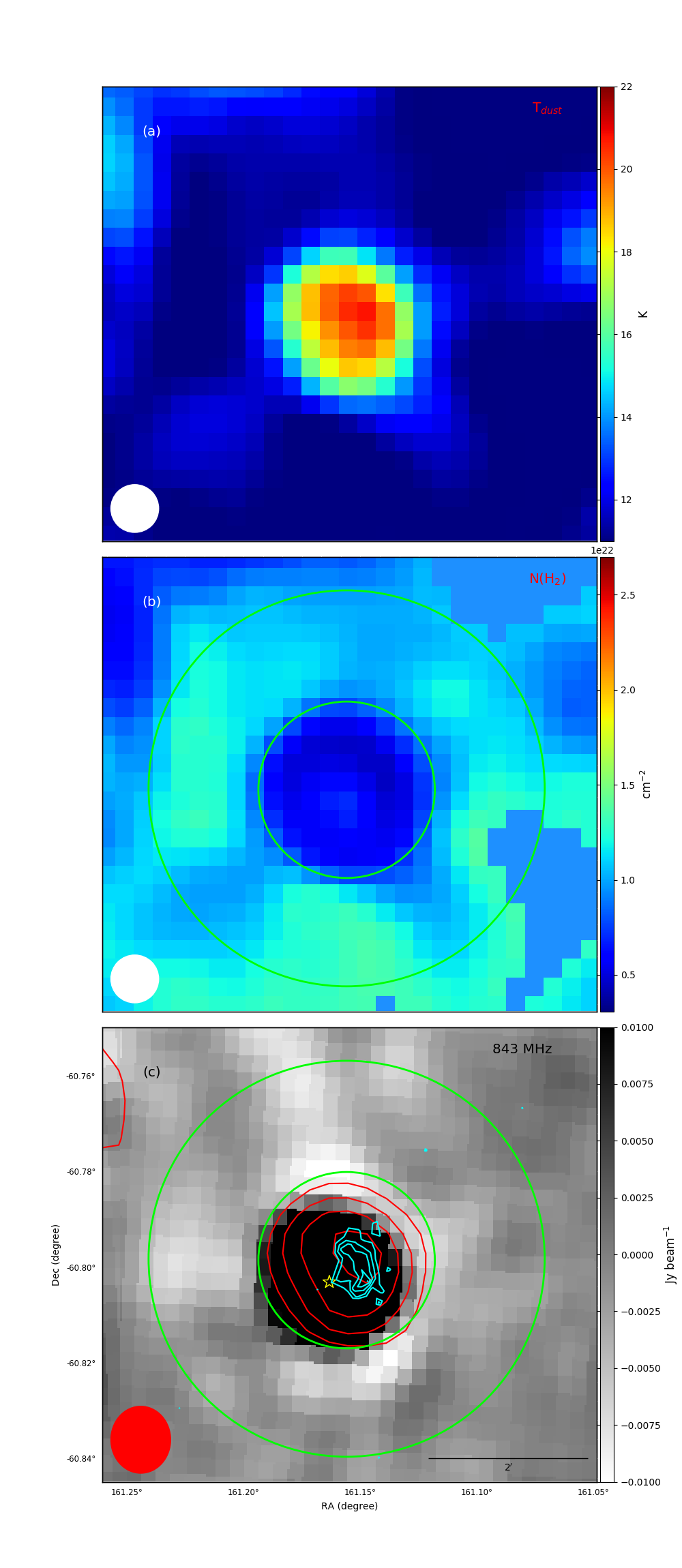}}
\caption{(a) Dust temperature map, and (b) column density map around IRAS\,10427-6032 for a 5\farcm5 $\times$5\farcm5 field derived from {\it Herschel} images in colour scale. We note some of the pixels of these images had NaN values (owing to bad pixel values in the original {\it Herschel} images). These pixels have been replaced with the values of the nearby pixels. (c) The 843 MHz radio continuum map, tracing the ionized gas around IRAS 10427-6032,
with overlaid dust temperature contours at 15,17,19 and 21 K. The green circles represent the approximate inner and outer boundaries of high column density
regions around the \ion{H}{II} region. The 870 $\mu$m  emission is shown in cyan contours.} 
\end{figure}

Figure~9 shows the low-resolution ($\sim$37${\arcsec}$), beam-averaged, dust temperature  and dust column density maps,  
and their correlations with the ionized gas over 5\farcm5 $\times$5\farcm5  
area centered on IRAS object. As can be seen from Figure~9a, though the temperature shows a distribution between 10 to 21 K, it is higher near the infrared cluster, peaking at $\sim$21 K. 
In contrast, the column density map (Figure~9b) shows, in general, a low column density towards the cluster center with an average value of 
$\sim$0.6 $\times$ 10$^{22}$ cm$^{-2}$, whereas relatively higher in the outskirts of the cluster, particularly within the annular area marked on the figure. The average value in the annular area is $\sim$1.3 $\times$ 10$^{22}$ cm$^{-2}$.  Figure~9c shows
the low-resolution ($\sim$43${\arcsec}$ $\times$ 49${\arcsec}$)  radio continuum view of the IRAS 10427-6032 region at 843 MHz. As can be seen, 
the radio emission is stronger at the center of the image 
corresponding to the location of the massive star (marked with a star symbol) and the cluster. The contours 
overlaid on the 843 MHz map are from the temperature map and are at 15, 17, 19 and 21 K.  The average temperature at the outskirts of the cluster, particularly in the annular area, is $\sim$10K.
Overall the 843 MHz and temperature maps show strong correlation, i.e., the warmer zone corresponds to the zone of stronger free-free emission implying that the relative high temperature observed at the cluster location is primarily due to the radiation from the members of the cluster. 
Though the {\it Herschel} maps are of low-resolution, yet a 
careful look seems to indicate a possible presence of a temperature gradient with temperature decreasing from northwest to southwest, 
consistent with the broken morphology of the \ion{H}{II} region, observed in the high-resolution optical and infra-red
images. These signatures suggest that the \ion{H}{II} region is possibly in its early phase of ``champagne-flow'' \citep{ten79}.

We estimate the total gaseous mass (M$_{gas}$) associated with the \ion{H}{II} region using the integrated column density over the size of the \ion{H}{II} region using the following equation: 
\begin{equation}
 M = \mu {\mathrm m_{H}} A_{\mathrm pix} \Sigma {\mathrm H_{2}} \, \label{eq:mass}
\end{equation}

where $\mu$ is the mean molecular weight, ${\mathrm m_{H}}$ is the mass of the hydrogen atom, $\Sigma$H$_2$ is the summed H$_2$ column density, and $A_{\mathrm pix}$ is the area of a pixel in cm$^{-2}$ at the distance of the region. The resultant mass is $\sim$ 220 M$_{\odot}$.

\subsection{Star formation}
Feedback from massive stars plays a critical role in the star formation processes and evolution of molecular clouds. In particular, expanding \ion{H}{II} 
regions may have a positive effect on star formation, i.e. they can 
trigger new  generation of star formation in molecular clouds. From a theoretical point of view, two main triggering mechanisms have been suggested: C\&C and RDI. In the C\&C process, when an  \ion{H}{II} region expands in a homogeneous medium, it 
sweeps the nearby ISM  into a dense shell. If the expansion of the \ion{H}{II} region continues for long enough, the surface density of 
the shell increases to the point where the shell becomes self-gravitating and fragments leading to the formation of 
massive condensations that are potential sites for subsequent star formation \citep{elmegreen77}. In the C\&C process as shown in simulations  \citet{whi94,dal07},
evenly spaced massive fragments are expected around the \ion{H}{II} regions \citep[e.g.][]{zav06,samal14,liu16}.  However, 
molecular clouds are often fractal and clumpy. Thus the clumpiness of the dense shell (or dense condensations in the shell) can be attributed to density 
structures in the fractal  molecular cloud into which the  \ion{H}{II} 
region \citep[see discussions in][]{wal13} expands. In RDI, an expanding \ion{H}{II} region overruns a pre-existing cloud, it drives an ionization front and a shock wave into the cloud. As a consequence, the inner parts of the 
cloud are compressed, and may become gravitationally unstable, collapsing to form new stars \citep{bertoldi89,bis11}.  Protruding structures 
(e.g., elephant trunks or  bright rimmed clouds) found at the edges of the \ion{H}{II} regions with YSOs or cores inside, are often considered as the signature of the RDI process \citep[e.g.][]{mor08,cha11a,cha11b,pan14}.

\citet{wal15} performed smoothed particle hydrodynamics simulations of \ion{H}{II} regions expanding into fractal molecular clouds and suggested that in a clumpy medium, a hybrid form of triggering, which combines elements of C\&C and RDI, should be more appropriate \citep[e.g.][]{jose13}. 
They found, in a fractal medium, during the expansion of the \ion{H}{II} region and the collection of the dense shell,
the pre-existing density structures are enhanced and lead to a clumpy distribution within the shell. 
The masses and locations of the clumps depend on the fractal density structure of the molecular cloud. Subsequently, the clumps grow in mass,
and at the same time they are overrun and compressed by the \ion{H}{II} region, until they become gravitationally unstable
and collapse to form new stars. 

As discussed in \S{3.3}, the annular area around the \ion{H}{II} region represents the location of
higher column densities. The average column density within the  annular area is approximately higher by a factor of two than the average column density within the \ion{H}{II} region.  Though the resolution of the
{\it Herschel} images are not high enough to discuss the morphology of the dust around the compact \ion{H}{II} region in
finer details, in general the column density distribution around the  \ion{H}{II} region broadly represents the accumulated cold  
matter such as those observed at the borders of several Galactic \ion{H}{II} regions \citep[e.g][]{deharveng10}. 
Largely, it appears that the \ion{H}{II} region has accumulated some of the diffuse ISM  into a shell.

We find the observed column density in the shell is comparable to the column density required ($\sim$ 6 $\times$ 10$^{21}$ cm$^{-2}$) 
for fragmentation to happen through C\&C process \citep[see][]{whi94}. Thus the shell may be in its initial stage of fragmentation.
However, as discussed in Sect. 3.1.2,  a compact 870 $\mu$m ATLASGAL clump lies at the western side of the infrared bubble, and this is the only 
ATLASGAL clump observed around the  \ion{H}{II} region. We note ATLASGAL 870 $\mu$m images are more sensitive to dense cold gas than diffuse gas.
The clump lies $\sim$27 arcsec away from the massive star (see Fig. 9c) and protrudes into the ionization region. 
Majority of the YSO candidates identified in the region are found to be 
coincident with the ATLASGAL clump, indicating the star formation in the clump is more active compared to the other parts of the region.  Our results suggest, although the \ion{H}{II} region has collected  some of the cold ISM around its periphery (perhaps through  C\&C process),
the enhanced star formation observed at its western side is unlikely due 
to the  fragmentation of the collected material, 
rather seems due to the compression of a pre-existing clump.  The fact that the average age of seven YSO candidates with SED fits, 0.17 Myr, is smaller by a factor of 2--5 as compared to the dynamical age of the \ion{H}{II} region (0.30--0.95 Myr), supports the role of expanding \ion{H}{II} region in triggering star formation in the clump. To put 
the star formation scenario of the region on a firm footing though, 
a detailed velocity and age measurements of the point sources, and  kinematics of the cold gas are needed. 

\section{Conclusions}
We studied a 5$\arcmin \times$5$\arcmin$ region around a compact \ion{H}{II} region, IRAS\,10427-6032, using the near-IR data from the VCNS, archival data from {\it Spitzer}, WISE, ${\it Herschel}$, ATLASGAL, and MGPS-2. We identified YSO candidates of the region using a combination of near-IR and mid-IR data.  Our conservative criteria result in 5 Class I and 29 Class II YSO candidates.  The ratio ($\sim$6) of the Class II to Class I YSO candidates suggests that this is a young cluster.  We derived approximate physical parameters of seven YSO candidates with photometric information in 7 or more bands, by constructing SEDs.  The SED fits show that these YSO candidates are all intermediate mass with masses ranging from $\sim$2 to 5 M$_{\odot}$, and in early evolutionary stages with an average age $\sim$0.17 Myr. Whereas the brighter sources are found to lie along a 1 Myr reddened (A$_V$=5.0 mag) main-sequence isochrone, the low- and intermediate mass- YSO candidates are clustered around a 1 Myr and 2 Myr pre-main sequence isochrones.  The mass distribution of all YSO candidates based on the isochrone fitting ranges from 0.1 M$_{\odot}$ to 5 M$_{\odot}$. 

The 843 MHz radio continuum data shows a nearly spherical compact radio source. The linear dimension of the source, assuming the distance to the region 2.3 kpc, is $\sim$1.2 pc. This implies that IRAS\,10427-6032 is a compact \ion{H}{II} region.  The Lyman continuum luminosity of the source, $10^{46.9}$ photons $\mathrm{s^{-1}}$,  suggests a ZAMS spectral type of the ionizing source to be B0.5--B1 or earlier assuming a single responsible source.  The candidate massive star is found at $\sim$9$\arcsec$ from the IRAS position and correlates well with the ionized emission.  Its expected spectral type based on the SED fit, B0--B0.5, also matches with the Lyman continuum luminosity derived from the radio continuum data.  The dynamical age of the \ion{H}{II} region is estimated to range between 0.30--0.95 Myr.

We present low-resolution ($\sim$ 37${\arcsec}$), beam-averaged, dust temperature  and dust column density maps generated using the ${\it Herschel}$ data.  The temperature distribution is found to vary between 10 to 21 K with  higher temperature peaking at $\sim$21 K near the location of the infrared cluster, and an average value of $\sim$15.5 K, away from the cluster. In contrast, the column density map shows a low column density towards cluster center with an average value of $\sim$0.6 $\times$ 10$^{22}$ cm$^{-2}$, whereas a relatively higher, $\sim$1.3 $\times$ 10$^{22}$ cm$^{-2}$, in the annular area around the \ion{H}{II} region.  This annular region likely represents the accumulated cold matter around the \ion{H}{II} region which is in its initial stage of fragmentation.

IRAS\,10427-6032 shows a broken bubble morphology in mid-IR images. The bubble of $\sim$1\farcm6 diameter has approximately two-third of its western rim intact and about one-third of the eastern side open.  The presence of temperature gradient, in which the temperature is seen to decrease from northwest to southwest in the temperature profile constructed using the ${\it Herschel}$ data, is consistent with the broken morphology of the \ion{H}{II} region, hinting that the \ion{H}{II} region is possibly in its early phase of ``champagne-flow''. 

The 870 $\mu$m ATLASGAL contours are along the western rim of the bubble, and show some amount of protruding in the ionized region.  From the spatial correlation it appears to be an interacting cold dust condensation.  Majority of the identified YSO candidates are found to be coincident with the sub-mm contours and are either found to lie along the bubble rim or into the bubble interior adjacent to the western rim.  Two of our five Class I YSO candidates are found in the dense shell surrounding the \ion{H}{II} region, one on the bubble rim, whereas the remaining two are found in the bubble interiors.  The spatial correlation of YSO candidates with the clump, and the greater dynamical age of the \ion{H}{II} region, by a factor $\sim$2--5, than the average age of the YSO candidates, indicate that the enhanced star formation on the western rim of the \ion{H}{II} region could be due to compression of the pre-existing clump. Spectroscopic information of the ionizing star of the \ion{H}{II} region and YSO candidates on the border, is necessary to strengthen the hypothesis of triggering in this star forming region. 

\section*{Acknowledgements}
This research has made use of the VizieR catalogue access tool, CDS, Strasbourg, France. The original description of the VizieR service was published in A\&AS 143, 23.  This work makes use of the archival images and catalogues from Deep Glimpse Survey of the {\it Spitzer} Science Telescope.  This publication makes use of data products from the Wide-field Infrared Survey Explorer, which is a joint project of the University of California, Los Angeles, and the Jet Propulsion Laboratory/California Institute of Technology, funded by the National Aeronautics and Space Administration.  This work also makes use of the archival data of 843 MHz radio continuum images observed under the second epoch of Molonglo Galactic Plane Survey.  This research makes use of the archival data from the ${\it Herschel}$ far-IR telescope.  The ATLASGAL project is a collaboration between the Max-Planck-Gesellschaft, the European Southern Observatory (ESO) and the Universidad de Chile. It includes projects E-181.C-0885, E-078.F-9040(A), M-079.C-9501(A), M-081.C-9501(A) plus Chilean data. This work makes use of the Python based SED fitting tool of \citet{robitallie06}.






\end{document}